\documentclass{emulateapj}

\begin{document}

\shortauthors{Moffat and Toth} \shorttitle{Testing modified gravity with globular clusters}

\title{Testing modified gravity with globular cluster velocity dispersions}

\author{J. W. Moffat\altaffilmark{1,2} and V. T. Toth\altaffilmark{1}}

\altaffiltext{1}{Perimeter Institute, 31 Caroline St North, Waterloo, Ontario N2L 2Y5, Canada}
\altaffiltext{2}{Department of Physics, University of Waterloo, Waterloo, Ontario N2L 3G1, Canada}

\begin{abstract}
Globular clusters (GCs) in the Milky Way have characteristic velocity dispersions that are consistent with the predictions of Newtonian gravity, and may be at odds with Modified Newtonian Dynamics (MOND). We discuss a modified gravity (MOG) theory that successfully predicts galaxy rotation curves, galaxy cluster masses and velocity dispersions, lensing, and cosmological observations, yet produces predictions consistent with Newtonian theory for smaller systems, such as GCs. MOG produces velocity dispersion predictions for GCs that are independent of the distance from the galactic center, which may not be the case for MOND. New observations of distant GCs may produce strong criteria that can be used to distinguish between competing gravitational theories.
\end{abstract}


\section{Introduction}

Modified Gravity (MOG, \cite{Moffat2005,Moffat2006a,Moffat2006b,Moffat2007e}) is a fully covariant theory of gravity that is based on postulating the existence of a massive vector field, $\phi_\mu$. The choice of a massive vector field is motivated by our desire to introduce a {\em repulsive} modification of the law of gravitation at short range. The vector field is coupled universally to matter. The theory yields a Yukawa-like modification of gravity with three constants: in addition to the gravitational constant $G$, we must also consider the coupling constant $\omega$ that determines the coupling strength between the $\phi_\mu$ field and matter, and a further constant $\mu$ that arises as a result of considering a vector field of non-zero mass, and controls the coupling range. In the most general case, these constants must be allowed to run. An approximate solution of the MOG field equations \citep{Moffat2007e} allows us to compute the values of $\mu$ and $\omega$ as functions of the source mass.

MOG was used successfully to describe observational phenomena on astrophysical and cosmological scales without resorting to dark matter. The theory correctly predicts galaxy rotation curves \citep{Brownstein2006a,Moffat2007e}, the mass and thermal profiles of clusters of galaxies \citep{Brownstein2006b,Moffat2007e}, the merging of the two clusters (Bullet Cluster, \cite{Brownstein2007}), the acoustical peaks of the cosmic microwave background \citep{Moffat2006b,Moffat2007d}, the velocity dispersion of satellite galaxies \citep{Moffat2007}, the mass power spectrum and the luminosity-distance relationship of distant Type Ia supernovae \citep{Moffat2007d}.

Globular clusters (GCs) in the Milky Way provide a particularly interesting case for testing alternate gravity theories \citep{BGK2005,Scarpa2007}, such as MOG or Milgrom's Modified Newtonian Dynamics (MOND, \cite{Milgrom1983,Bekenstein2004}).

GCs at different distances from the galactic center experience galactic gravity at varying strengths. The internal gravitational field of a GC also varies depending on the mass (typically, 10$^4$--10$^6~M_\odot$) and size (typically, a few pc to a few ten pc in diameter) of the GC in question. Using the characteristic acceleration ($a_0=1.2\times 10^{-10}$~m/s$^2$) of MOND, for example, we find GCs that experience either galactic or internal acceleration above, or below this value. MOND predicts velocity dispersions different from the Newtonian prediction for a GC whose stars experience a combined acceleration less than $a_0$.

On the other hand, MOG predicts identical velocity dispersions for GCs of the same size and composition, regardless of their distance from the galactic center. For this reason, GCs provide a unique test by which different gravitational theories can be compared.

\section{Theory}

The bulk properties of GCs, with the possible exception of their innermost regions, can be modeled using the collisionless Boltzmann equation \citep{BT1987}, from which the statistical properties of the velocity distribution of stars can be derived. In particular, one can derive a formulation for the velocity dispersion tensor that, in the isotropic, non-rotating case, reduces to a scalar quantity. This quantity can be determined using the appropriate Jeans equation. For this calculation, one requires knowledge of the distribution function (DF) that determines the number of stars in a given region of space, and the gravitational potential. We begin our discussion with the latter.

\subsection{Modified Gravity}

Our modified gravity (MOG) theory predicts a Yukawa-like modification of Newton's law of acceleration \citep{Moffat2006a,Moffat2007e}, in the form
\begin{equation}
a_\mathrm{MOG}=-\frac{G_NM}{r^2}(1+\alpha(1-(1+\mu r)e^{-\mu r})),
\end{equation}
where $G_N$ is Newton's gravitational constant.

In accordance with our recent results \citep{Moffat2007e}, the parameters $\alpha$ and $\mu$ can now be {\em predicted}:
\begin{eqnarray}
\alpha&=&\frac{M}{(\sqrt{M}+C_1')^2}\left(\frac{G_\infty}{G_N}-1\right),\\
\mu&=&\frac{C_2'}{\sqrt{M}},
\end{eqnarray}
where
\begin{eqnarray}
G_\infty&\simeq&20G_N,\\
C_1'&\simeq&25000~M_\odot^{1/2},\\
C_2'&\simeq&6250~M_\odot^{1/2}\mathrm{kpc}^{-1}.
\end{eqnarray}

For even a large GC, with mass exceeding $10^6~M_\odot$, the predicted values are
\begin{eqnarray}
\alpha&\simeq&0.03,\\
\mu&\simeq&(160~\mathrm{pc})^{-1}.
\end{eqnarray}

Given the smallness of $\alpha$ and the fact that $\mu^{-1}$ is much larger than the GC radius, it is clear that our theory predicts Newtonian behavior for such GCs:
\begin{equation}
a_\mathrm{MOG}\simeq a_\mathrm{Newton}=-\frac{G_NM}{r^2}.
\end{equation}
For smaller GCs, the predictions are even closer to Newtonian values.

In contrast, the MOND acceleration $a_{\rm MOND}$ is given by the solution of the non-linear equation
\begin{equation}
a_{\rm MOND}\mu\left(\frac{|a_{\rm MOND}|}{a_0}\right)=-\frac{G_NM}{r^2},
\end{equation}
where $a_0=1.2\times 10^{-8}{\rm cm}\: {\rm sec}^{-2}$. The form of the function $\mu(x)$ originally proposed by \cite{Milgrom1983} is given by $\mu(x)=x/\sqrt{1+x^2}$; however, better fits and better asymptotic behavior are achieved using $\mu(x)=x/(1+x)$~\citep{Famaey2005}, which yields the acceleration function
\begin{equation}
a_\mathrm{MOND}=-\frac{G_NM}{r^2}\left(\frac{1}{2}+\sqrt{\frac{1}{4}+\frac{a_0r^2}{G_NM}}\right).
\end{equation}
Regardless of the form of $\mu(x)$ chosen, when the combined acceleration experienced by stars in a GC is below $a_0$, MOND predicts dynamical behavior that is markedly different from the Newtonian prediction.

\subsection{The Jeans equation}

In the spherically symmetric, non-rotating case the Jeans equation for the velocity dispersion $\sigma(r)$ takes the following form (see Eq. 4-64a in \cite{BT1987}):
\begin{equation}
\frac{\partial(\nu\sigma^2)}{\partial r}+\nu\frac{\partial\Phi}{\partial r}=0,
\label{eq:Jeans}
\end{equation}
where $r$ is the radial distance from the GC center, $\nu(r)$ is the number density distribution function, and $\Phi(r)$ is the gravitational potential. If $\nu(r)$ and $\Phi(r)$ are known, the velocity dispersion can be obtained by direct integration. Using $a(r)=\partial\Phi/\partial r$ and utilizing the fact that $\lim\limits_{r\rightarrow\infty}\sigma^2(r)=0$, we get
\begin{equation}
\sigma^2(r)=\frac{1}{\nu}\int\limits_r^\infty\nu a(r')~dr'.
\label{eq:sigma}
\end{equation}

The observed velocity dispersion of GCs is a function not of the actual radial distance $r$ but the projected distance $R$ between the GC center and the star being observed. The velocity dispersion $\sigma_\mathrm{LOS}(R)$ of stars observed along the line-of-sight (LOS) at projected distance $R$ from the GC center is related to $\sigma(r)$ as
\begin{equation}
\sigma_\mathrm{LOS}^2(R)=\frac{\int_0^\infty\sigma^2(y)\nu(y)~dy}{\int_0^\infty\nu(y)~dy}
\label{eq:LOSy}
\end{equation}
where
\begin{equation}
y^2=r^2-R^2,
\end{equation}
as can be verified by simple geometric reasoning. Eliminating $y$, we can rewrite (\ref{eq:LOSy}) as
\begin{equation}
\sigma_\mathrm{LOS}^2(R)=\frac{\int_R^\infty r\sigma^2(r)\nu(r)/\sqrt{r^2-R^2}~dr}{\int_R^\infty r\nu(r)/\sqrt{r^2-R^2}~dr}.
\label{eq:LOS}
\end{equation}

\subsection{Density Distribution}

Several models exist that can mimic the density distribution of a spherically symmetric set of stars. One particularly simple model is that of \cite{HERN1990}, which models the number density of stars as a function of radius as
\begin{equation}
\nu_\mathrm{Hernquist}(r)=\frac{Nr_0}{2\pi r(r+r_0)^3},
\end{equation}
where $N$ is the total number of stars, and $r_0$ is a characteristic radius.

Another, similar model is that of \cite{JAFFE1983}:
\begin{equation}
\nu_\mathrm{Jaffe}(r)=\frac{Nr_0}{4\pi r^2(r+r_0)^2}.
\end{equation}

Without benefiting from a photometric profile of the globular cluster under investigation, there are no {\em a priori} reasons to prefer one model over another. We found that the choice of model does not have a significant impact on the conclusions we present; hereinafter, we shall be using Hernquist's model consistently, but we note that similar results are obtained using alternate number density distribution functions.

\section{Observations and predictions}

\begin{figure*}[th]
\begin{center}
\hskip -6pt
\begin{minipage}[b]{.32\linewidth}
\centering \plotone{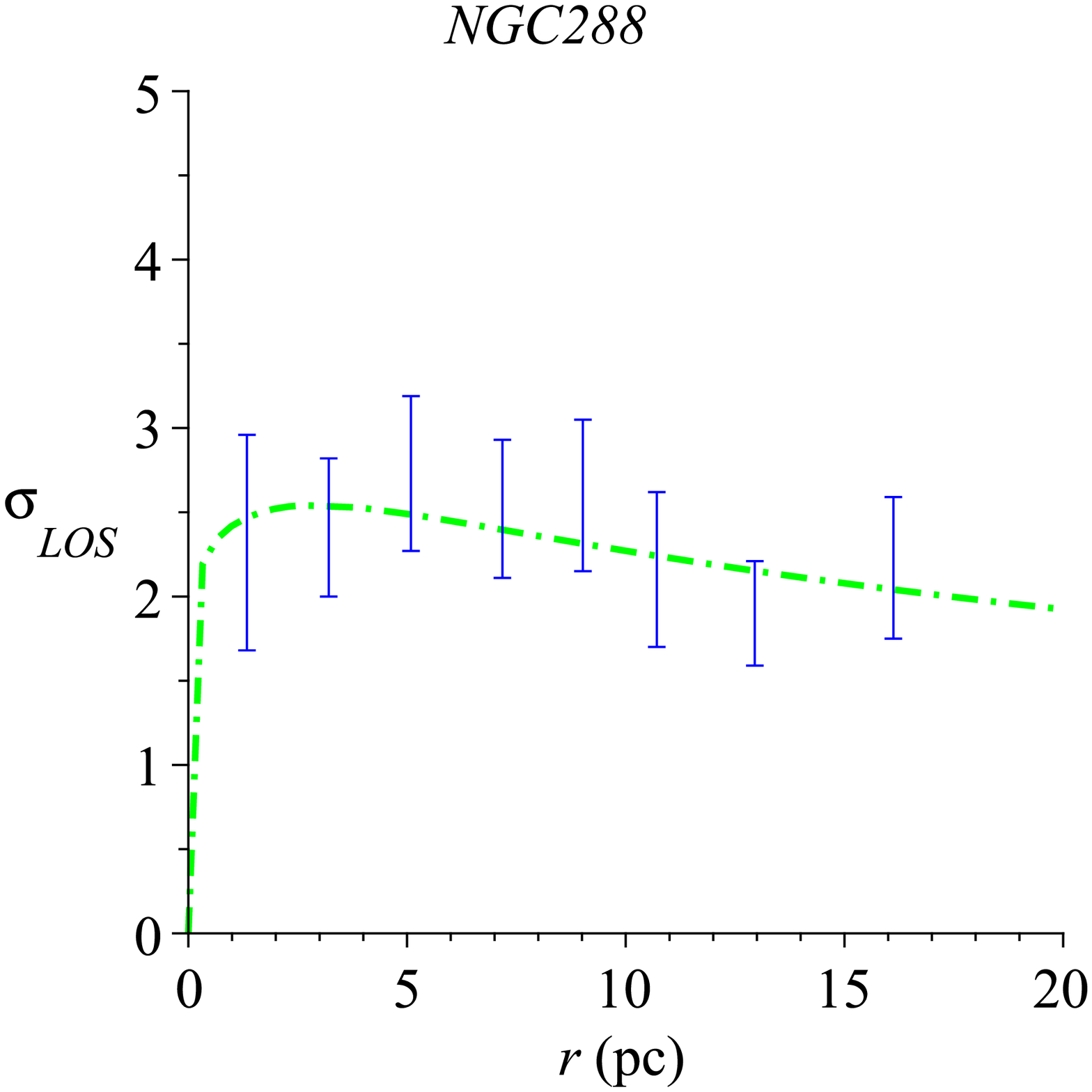}
\end{minipage}
\hskip 0.001\linewidth
\begin{minipage}[b]{.32\linewidth}
\centering \plotone{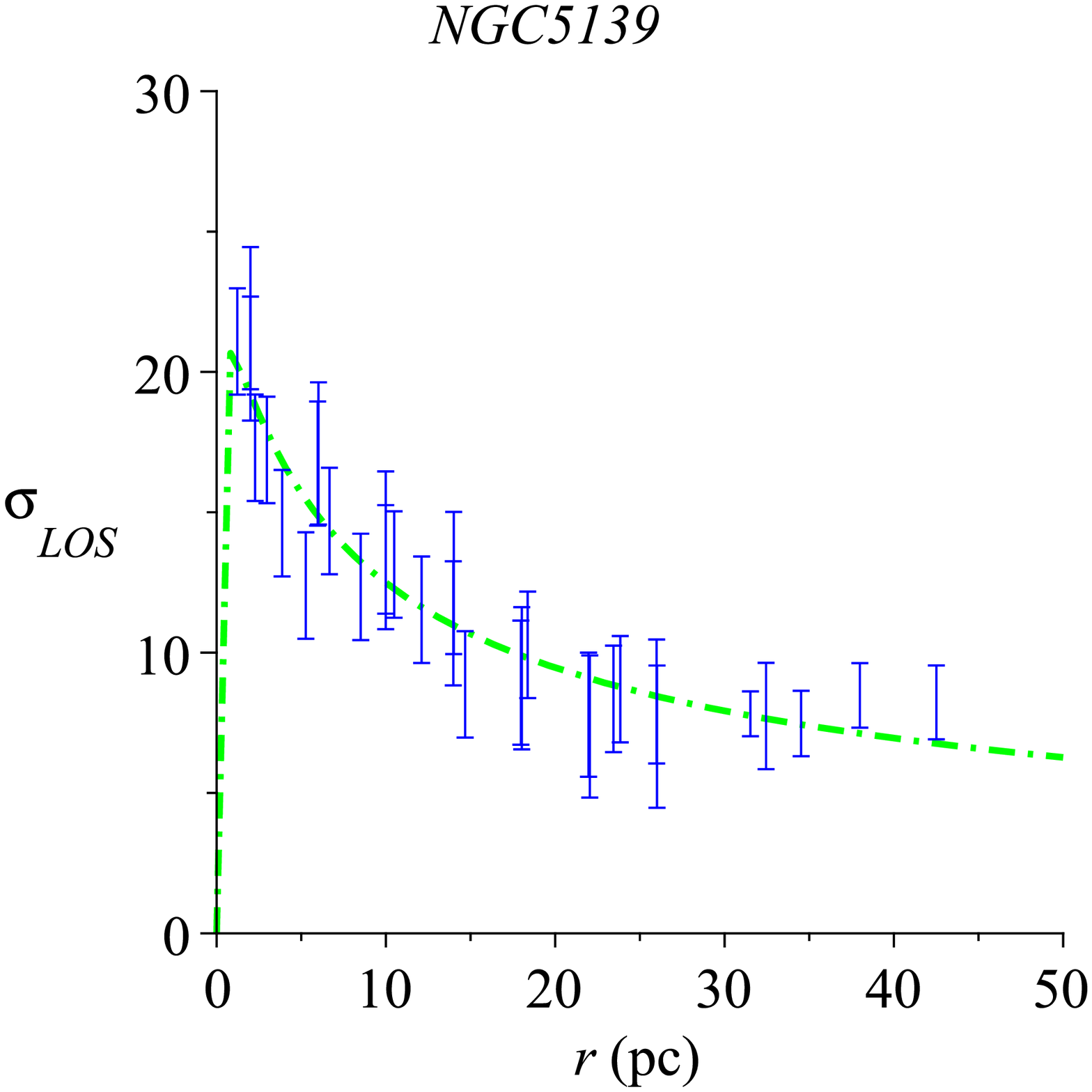}
\end{minipage}
\hskip 0.001\linewidth
\begin{minipage}[b]{.32\linewidth}
\centering \plotone{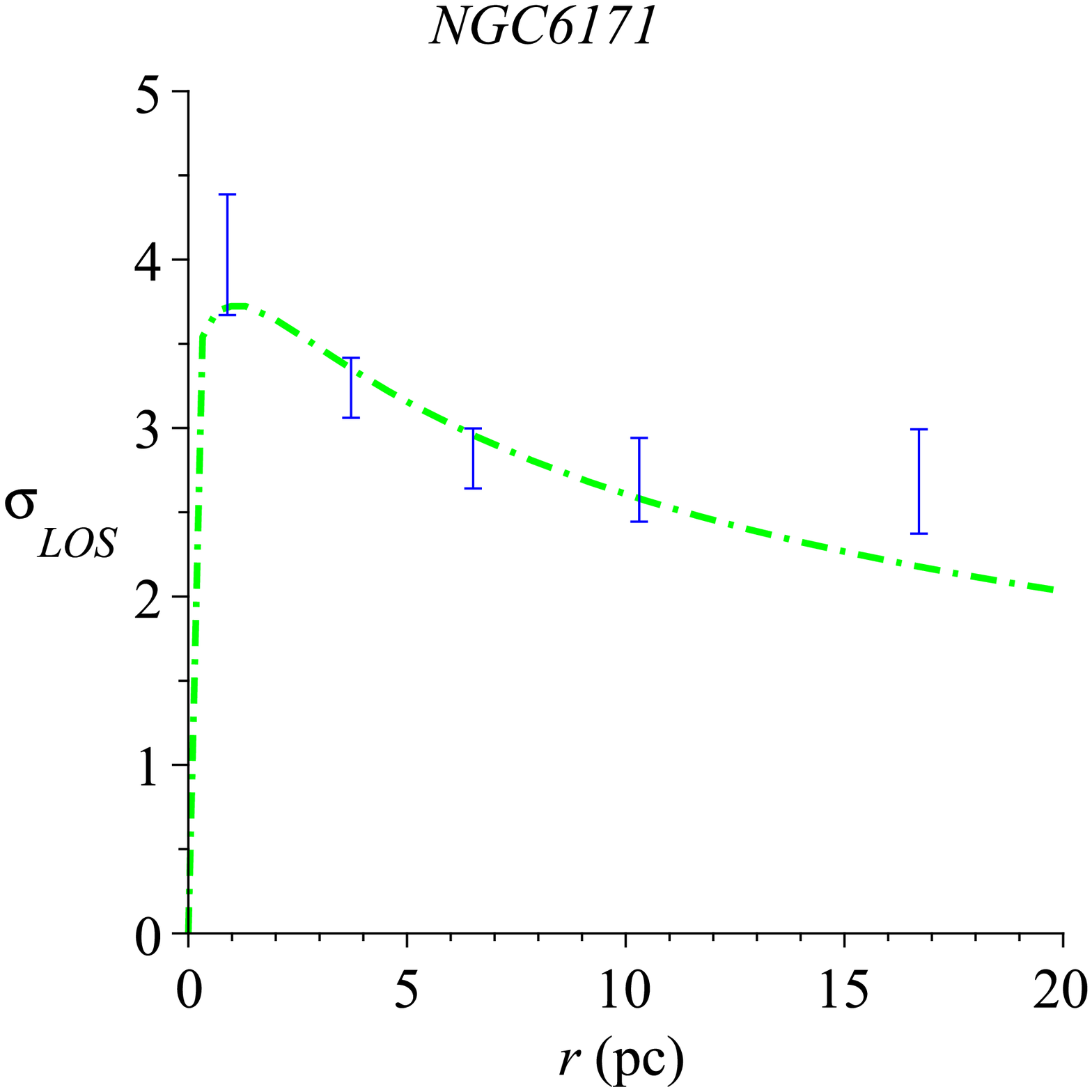}
\end{minipage}
\par
\hskip -6pt
\begin{minipage}[b]{.32\linewidth}
\centering \plotone{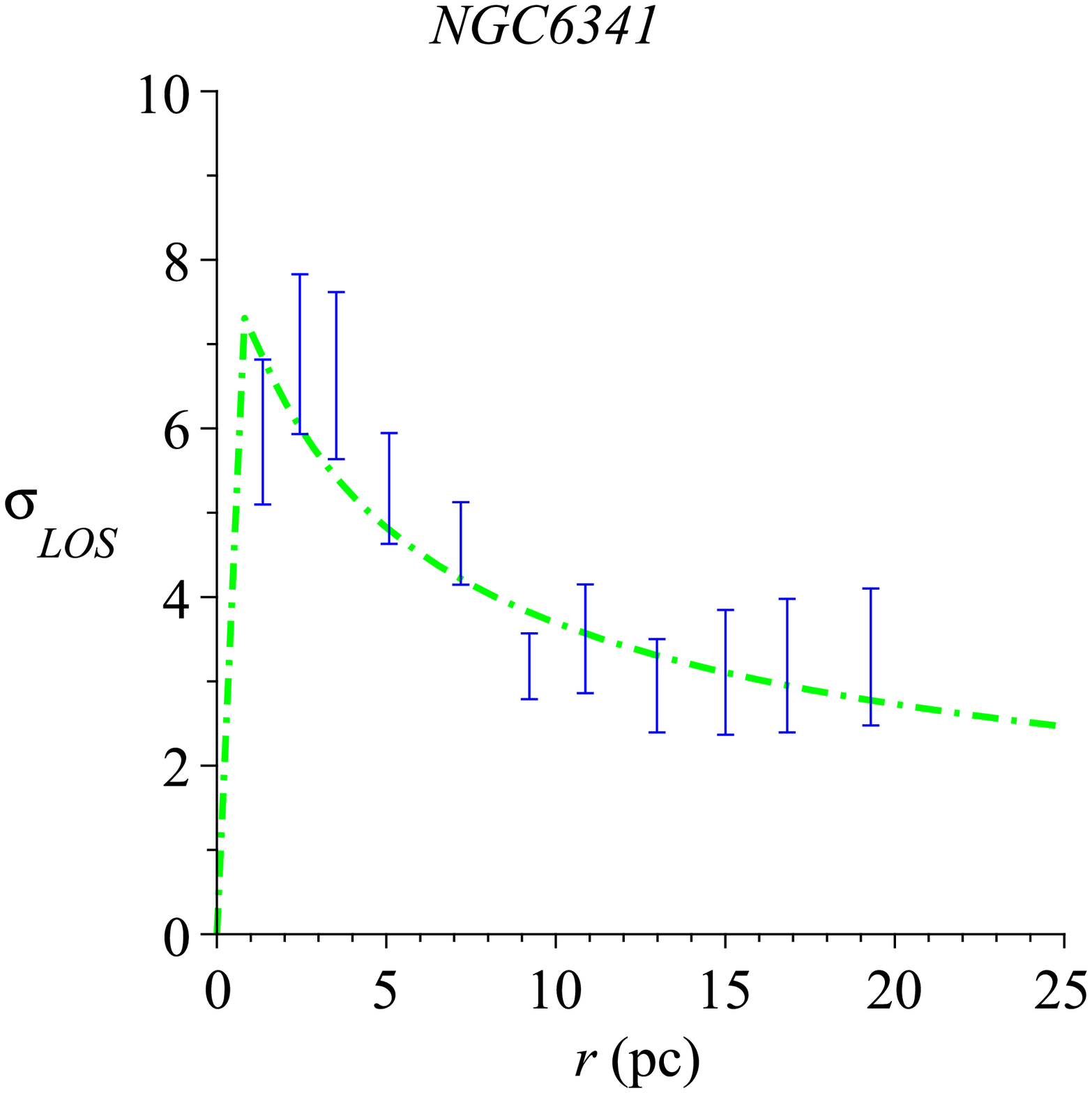}
\end{minipage}
\hskip 0.001\linewidth
\begin{minipage}[b]{.32\linewidth}
\centering \plotone{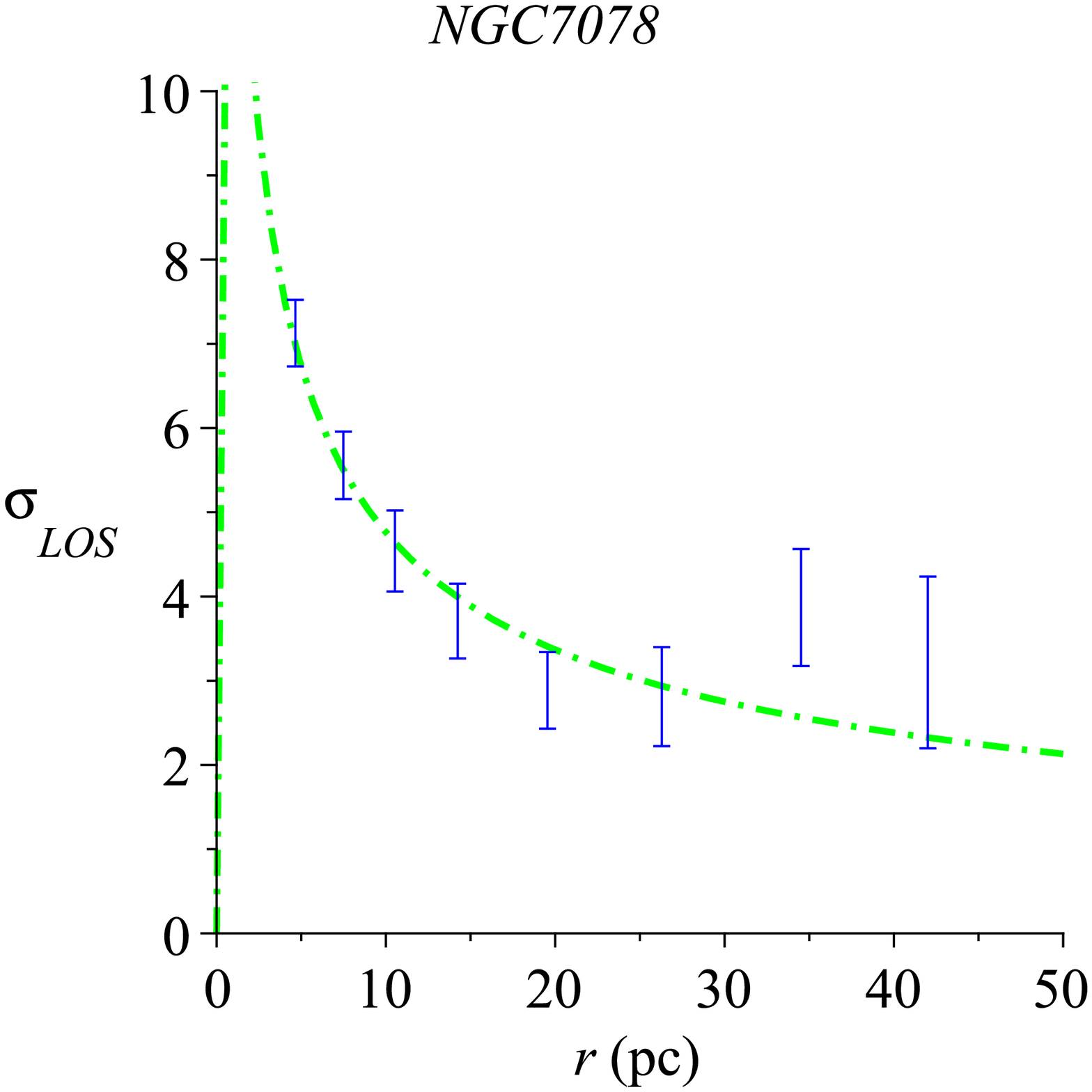}
\end{minipage}
\hskip 0.001\linewidth
\begin{minipage}[b]{.32\linewidth}
\centering \plotone{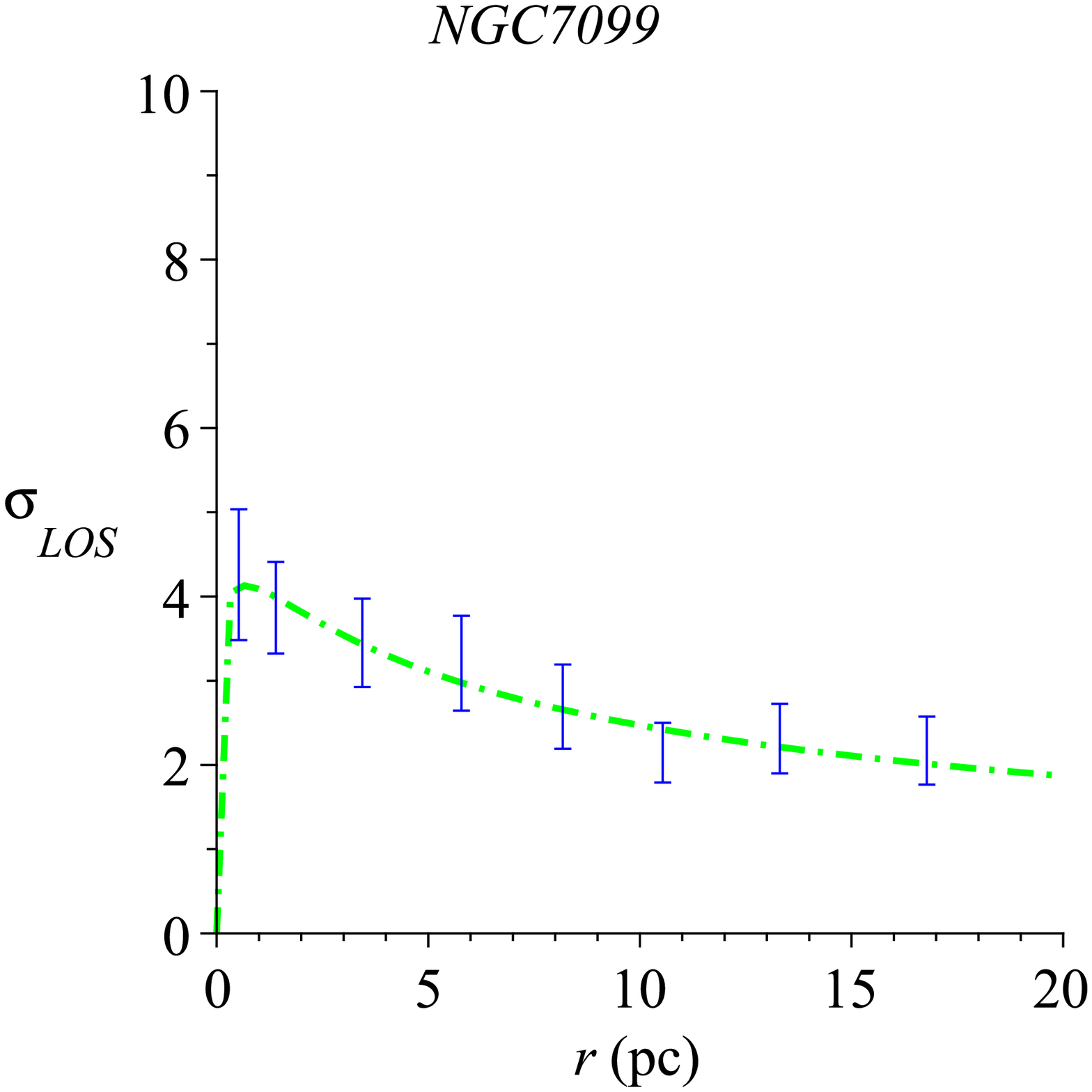}
\end{minipage}
\end{center}
\vskip -12pt
\caption{Fitting velocity dispersions obtained from the Jeans equation to globular cluster data (blue error bars from \cite{Scarpa2007}), using the Hernquist model and MOG or Newtonian gravity (dash-dot green line).}
\label{fig:clusters}
\end{figure*}

\begin{figure*}[th]
\hskip 0.02\linewidth
\begin{minipage}[b]{.4\linewidth}
\centering \includegraphics[width=\linewidth]{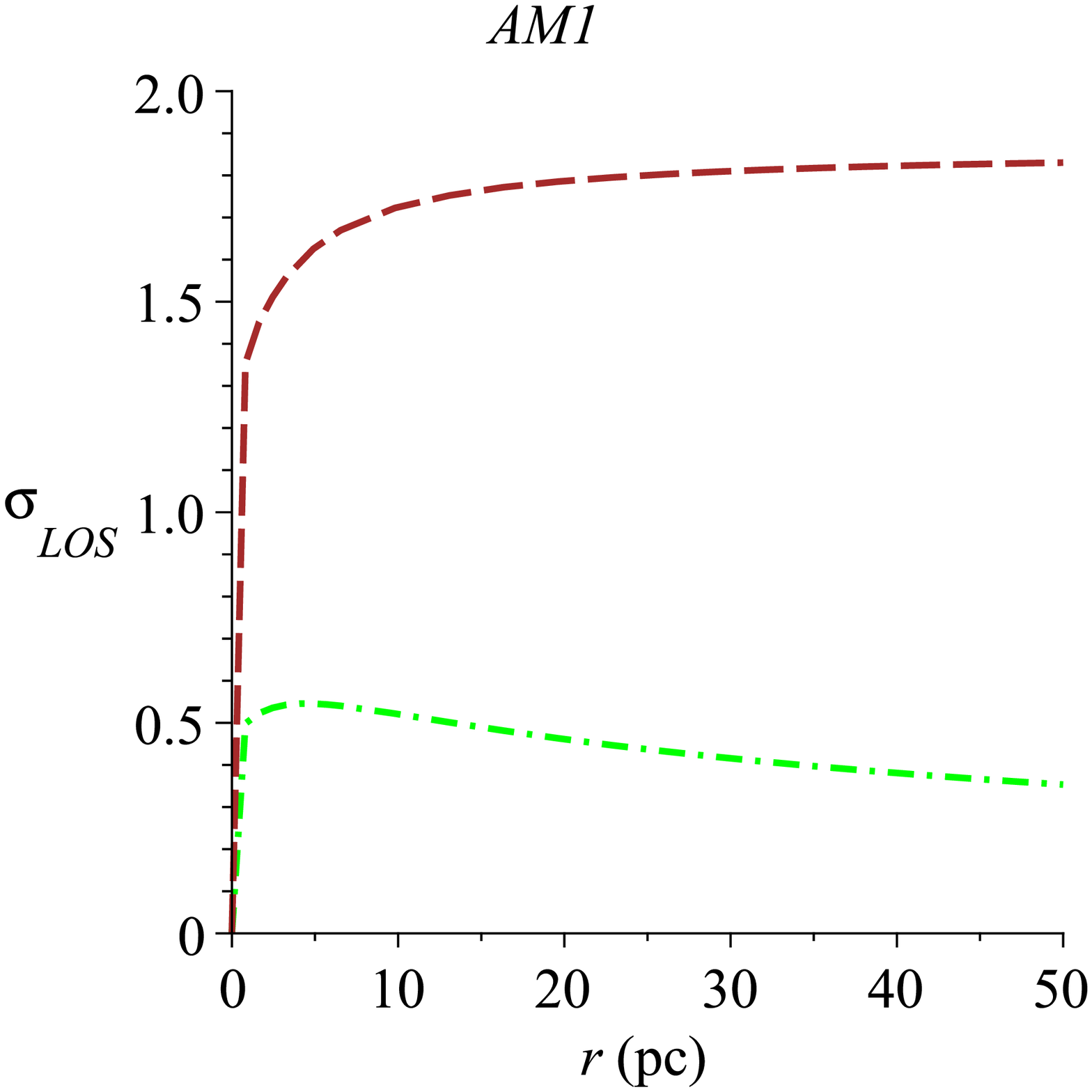}
\end{minipage}
\hskip 0.1\linewidth
\begin{minipage}[b]{.4\linewidth}
\centering \includegraphics[width=\linewidth]{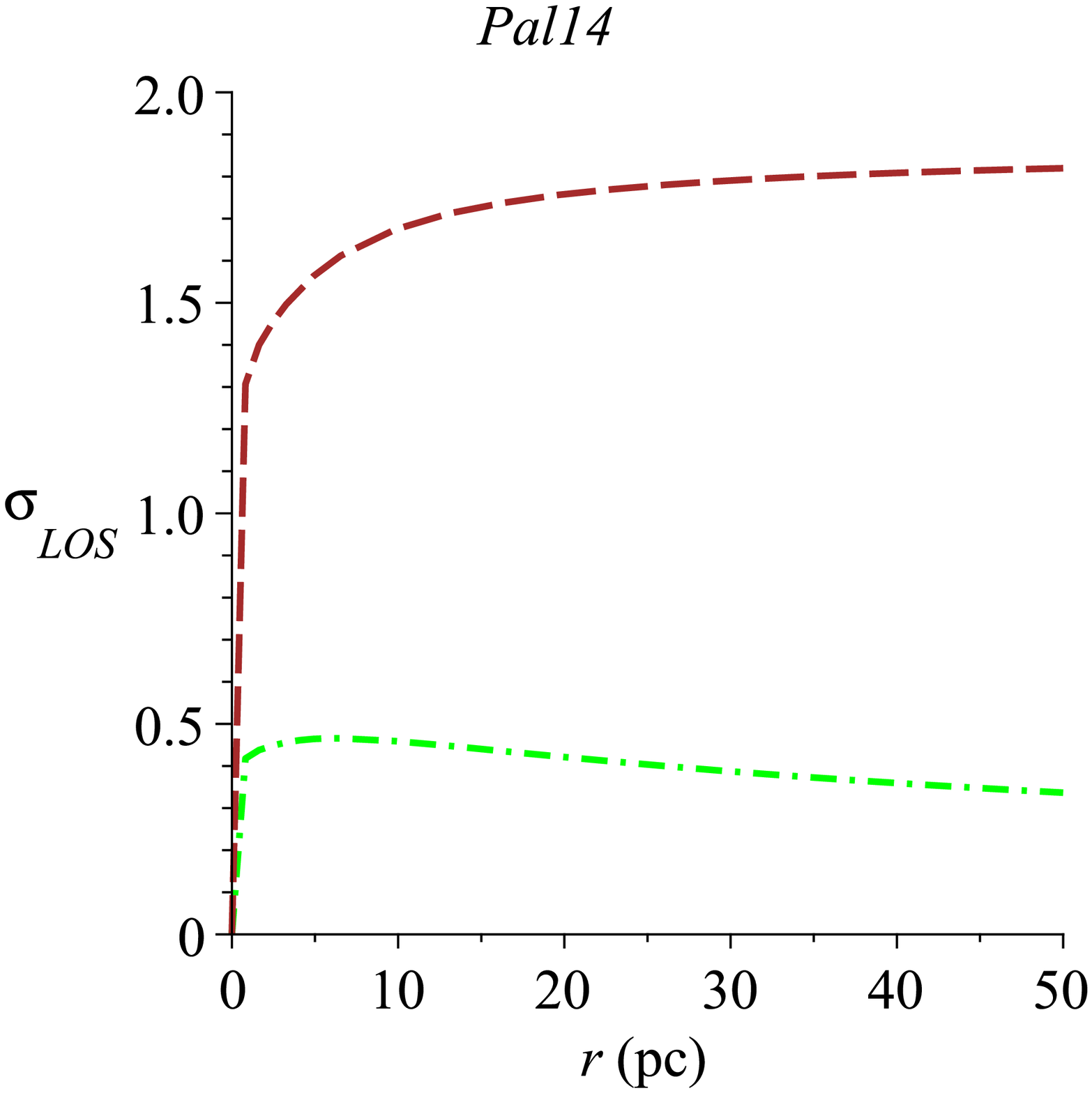}
\end{minipage}
\vskip -5pt
\caption{Predicted velocity dispersion curves for two distant GCs. Dash-dot line (green) is the prediction obtained using MOG or Newtonian gravity; dashed (brown) curve is the MOND prediction. In both cases, we used $M/L=2$ and we equated the parameter $r_0$ of the Hernquist model with the half-light radius.}
\label{fig:pred}
\end{figure*}

\begin{table}
\caption{Properties of GCs studied by \cite{Scarpa2007}. Data for AM 1 and Pal 14 are also included. The distance $R_g$ from the galactic center, luminosity $L_\odot$ in units of solar luminosity, and the half-light radius $r_h$ (pc) are shown \citep{Harris1996}. Mass-to-light ratios are estimated by fitting the velocity dispersion using the Hernquist model, except for AM 1 and Pal 14, for which $M/L=2$ was fixed.}\par
\begin{center}
\begin{tabular}{|l|c|c|c|c|}\hline\hline
Name&$R_g$~(kpc)&$L_\odot$&$r_h$~(pc)&$M_\odot/L_\odot$\\\hline
NGC288&~~7.4&$3.94\times 10^4$&~2.9&4.38\\
NGC5139&~~6.4&$1.04\times 10^6$&~6.4&2.79\\
NGC6171&~~3.3&$5.65\times 10^4$&~5.0&2.54\\
NGC6341&~~9.6&$1.51\times 10^5$&~2.6&1.50\\
NGC7078&~10.4&$3.70\times 10^5$&~3.2&0.85\\
NGC7099&~~7.1&$7.45\times 10^4$&~2.7&1.51\\
AM 1&123.2&$6.08\times 10^3$&17.7&2\\
Pal 14&~69.0&$6.19\times 10^3$&24.7&2\\\hline
\end{tabular}
\end{center}
\label{tb:GC}
\end{table}

Velocity dispersion data for several GCs were recently published by \cite{Scarpa2007}. We read velocity dispersion values and their standard deviations from Figures 1--2 and 4 of \cite{Scarpa2007}, for NGC 288, NGC 5139 ($\omega$ Centauri), NGC 6171 (M107), NGC 6341 (M92), NGC 7078 (M15), and NGC 7099 (M30). Some of the basic characteristics of these GCs are summarized in Table~\ref{tb:GC}.

Using the Hernquist distribution as the number density distribution function for a spherically symmetric cluster of stars with isotropic velocity dispersion, we obtained very good fits to the velocity dispersion data (Figure~\ref{fig:clusters}). These results also yield mass-to-light ratios ranging between $0.8<M/L<4.4$ (Table~\ref{tb:GC}), which are typical for globular clusters.

For these results, we used the Newtonian gravitational potential. These calculations are consistent with Newtonian theory, MOG (given the smallness of the predicted value of the MOG $\alpha$ parameter and the large size of the parameter $\mu^{-1}$), and also MOND, as the GCs in question are located relatively near the galactic center, and the galactic acceleration is always greater than $a_0$.

The possible presence of dark matter does not appreciable alter these results either. A typical dark matter density for the galactic halo is $\sim 7.8\times10^{-3}~M_\odot/\mathrm{pc}^3$ ($\simeq$0.3~GeV/cm$^3$; see \cite{Sumner2002}), a density that is much smaller than the globular cluster's stellar mass density.

The situation is different, however, in the case of MOND and globular clusters that are located a long distance away from the galactic center. To quote \cite{Milgrom1983}: ``We are then compelled to conclude that the internal dynamics of the open clusters embedded in the field of the Galaxy is different from that of a similar but isolated cluster.'' For instance, Pal 14, located at 69~kpc from the galactic center, would experience a galactic acceleration of $\sim 2.3\times 10^{-11}$~m/s$^2$, which is well within the MOND regime. As this is a low mass cluster of stars, its internal accelerations are also significantly below MOND's $a_0$, except perhaps in the innermost regions of the cluster.

Two distant clusters, AM 1 and Pal~14, are presently the subject of an observational project by Kroupa et~al. As the absolute luminosity of these GCs is known, using a (typical) value of $M/L\simeq 2$ we can obtain a crude estimate of their mass, allowing us to apply the Jeans equation and derive a velocity dispersion profile using both Newtonian and MOND gravity. These predictions are shown in Figure~\ref{fig:pred}.

\section{Discussion}

For globular clusters with a mass of a few times $10^6~M_\odot$ or less, MOG predicts little or no observable deviation from Newtonian gravity. This is verified by our demonstration that a simple model, using a spherically symmetric distribution and no velocity anisotropy, can easily reproduce the velocity dispersion profiles of several diverse globular clusters with varying mass.

The predictions of neither Newtonian gravity nor MOG depend on the distance from the galactic center. The same remains true when dark matter is considered; although the density of dark matter may be a function of distance from the galactic center, at predicted dark matter densities, the amount of dark matter contained within a GC does not noticeably alter the dynamics of the cluster.

The situation is different for MOND; for a low-mass cluster, internal accelerations are below the MOND threshold of $a_0\simeq 1.2\times 10^{-10}$~m/s$^2$, and if the cluster is far enough from the galactic center, its galactic acceleration is also below this value. For this reason, distant GCs may offer a unique method to distinguish observationally between MOG and MOND.

\acknowledgements
The research was partially supported by National Research Council of Canada. Research at the Perimeter Institute for Theoretical Physics is supported by the Government of Canada through NSERC and by the Province of Ontario through the Ministry of Research and Innovation (MRI). We thank Pavel Kroupa for helpful and stimulating correspondence.

\bibliographystyle{apj}

\end{document}